\documentclass[12pt]{article}
\usepackage{graphicx}
\begin{document}

\begin{center}

{\Large \bf Symmetries of Massive and Massless Neutrinos}

\vspace{2mm}
Y. S. Kim\\
Center for Fundamental Physics, University of Maryland,\\
College Park, Maryland 20742, U.S.A.\\
e-mail: yskim@umd.edu \\

\vspace{2mm}

\end{center}

\begin{abstract}
Wigner's little groups are subgroups of the Lorentz group dictating
the internal space-time symmetries of massive and massless particles.
These little groups are like O(3) and E(2) for massive and massless
particles respectively.  While the geometry of the O(3) symmetry is
familiar to us, the geometry of the flat plane cannot explain the
E(2)-like symmetry for massless particles.  However, the geometry
of a circular cylinder can explain the symmetry with the helicity
and gauge degrees of freedom.  It is shown further that the symmetry
of the massless particle can be obtained as a zero-mass limit of
O(3)-like symmetry for massive particles.  It is shown further that
the polarization of massless neutrinos is a consequence of gauge
invariance, while the symmetry of massive neutrinos is still like O(3).
\end{abstract}

\vspace{30mm}
\noindent  presented at the International Conference on Hadron
Structure and QCD: from Low to High Energies (Gatchina, Russia, 2016)

\newpage

\section{Introduction}\label{intro}
In his 1939 paper~\cite{wig39}, Wigner considered the subgroups of
the Lorentz group whose transformations leave the four-momentum of
a given particle invariant.  These subgroups are called Wigner's
little groups and dictate the internal space-time symmetries in
the Lorentz-covariant world.  He observed first that a massive
particle at rest has three rotational degree of freedom leading
to the concept of spin.  Thus the little group for the massive
particle is like O(3).

Wigner observed also that a massless particle cannot be brought to
its rest frame, but he showed that the little group for the massless
particle also has three degrees of freedom, and that this little
is locally isomorphic to the group E(2) or the two-dimensional
Euclidean group.  This means that the
generators of this little group share the same set of closed
commutation relations with that for two-dimensional Euclidean
group with one rotational and two translational degrees of freedom.

It is not difficult to associate the rotational degree of freedom
of E(2) to the helicity of the massless particle.  However, what
is the  physics of the those two translational degrees of freedom?
Wigner did not provide the answer to this question in his 1939
paper~\cite{wig39}.  Indeed, this question has a stormy history and
the issue was not completely settled until 1990~\cite{kiwi90jmp},
fifty one years after 1939.

In this report, it is noted first that the Lorentz group has six
generators. Among them, three of them generate the rotation subgroup.
It is also possible to construct three generators which constitute
a closed set of commutations relations identical to that for the E(2)
group.   However, it is also possible to construct the cylindrical
group with one rotational degree of freedom and two degrees freedom
both leading to up-down translational degrees freedom.  These two
translational degrees freedom correspond to one gauge degree of
freedom for the massless particle~\cite{kiwi87jmp}.

While the O(3)-like and E(2)-like little groups are different, it
is possible to derive the latter as a Lorentz-boosted O(3)-like
little group in the infinite-momentum limit. It is shown then
that the two rotational degrees of freedom perpendicular momentum
become one gauge degree of freedom~\cite{hks83pl}.

It is noted that the E(2)-like symmetry for the massless spin-1
particle leads to its helicity and gauge degree of freedom.  Likewise,
there is a gauge degree of freedom for the massless spin-1/2 particle.
However, the requirement of gauge invariance leads to the polarization
of massless neutrinos~\cite{hks82,bkn15,kmn16}.

In Sec.~\ref{little}, we introduce Wigner's little groups for massive
and massless particles.  In Sec.~\ref{o3e2}, it is shown that the
E(2)-like little group for massless particles can be obtained as the
infinite-momentum limit of the O(3)-like little group.  In
Sec.~\ref{spinhalf}, the same logic is developed for spin-half
particles.  It is shown that the polarization of massless neutrinos is
a consequence of gauge invariance.

\section{Wigner's little groups}\label{little}
If we use the four-vector convention $x^{\mu} = (x, y, z, t)$, the
generators of rotations around and boosts along the $z$ axis take the
form
\begin{equation}\label{eq001}
J_{3} = \pmatrix{0&-i&0&0\cr i&0&0&0\cr 0&0&0&0\cr 0&0&0&0} , \qquad
K_{3} = \pmatrix{0&0&0&0\cr 0&0&0&0 \cr 0&0&0&i \cr 0&0&i&0} ,
\end{equation}
respectively.  We can also write the four-by-four matrices for $J_{1}$
and $J_{2}$ for the rotations around the $x$ and $y$ directions, as
well as $K_{1}$ and $K_{2}$ for Lorentz boosts along the $x$ and $y$
directions respectively~\cite{bkn15}.  These six generators satisfy
the following set of commutation relations.
\begin{equation}\label{eq002}
\left[J_{i}, J_{j}\right] = i\epsilon_{ijk} J_{k}, \qquad
\left[J_{i}, K_{j}\right] = i\epsilon_{ijk} K_{k}, \qquad
\left[K_{i}, K_{j}\right] = -i\epsilon_{ijk} J_{k}.
\end{equation}
This closed set of commutation relations is called the Lie algebra of
the Lorentz group.   The three $J_{i}$ operators constitute a closed
subset of this Lie algebra.  Thus, the rotation group is a subgroup
of the Lorentz group.

In addition, Wigner in 1939~\cite{wig39} considered a group generated
by
\begin{equation}\label{eq005}
J_{3}, \qquad N_{1} = K_{1} - J_{2} ,\qquad N_{2} = K_{2} + J_{1} .
\end{equation}
These generators satisfy the closed set of commutation relations
\begin{equation}\label{eq200}
\left[N_{1}, N_{2}\right] = 0, \qquad
\left[J_{3}, N_{1}\right] = iN_{2}, \qquad
\left[J_{3}, N_{2}\right] = -iN_{1}.
\end{equation}

\begin{figure}
\centerline{\includegraphics[scale=1.6]{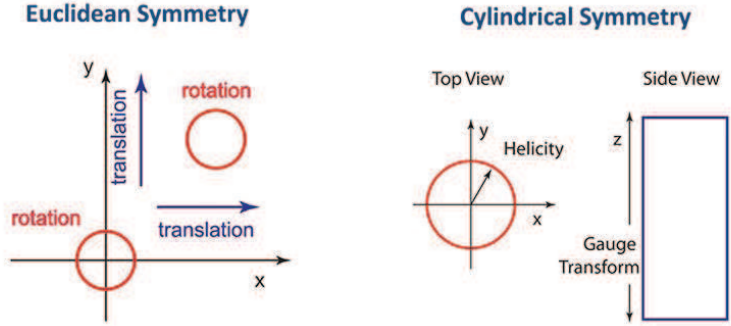}}
\caption{Transformations of the $E(2)$ group and the cylindrical
group.  They share the same Lie algebra, but only the cylindrical
group leads to a geometrical interpretation of the gauge
transformation. }\label{cylin}
\end{figure}

As Wigner observed in 1939~\cite{wig39}, this set of commutation
relations is just like that for the generators of the two-dimensional
Euclidean group with one rotation and two translation generators,
as illustrated in Fig.~\ref{cylin}.  However, the question is what
aspect of the massless particle can be explained in terms of this
two-dimensional geometry.

Indeed, this question has a stormy history, and was not answered
until 1987.  In their paper of 1987~\cite{kiwi87jmp}, Kim and Wigner
considered the surface of a circular cylinder as shown in
Fig.~\ref{cylin}.  For this cylinder, rotations are possible around
the $z$ axis.  It is also possible to make translations along the
$z$ axis as shown in Fig.~\ref{cylin}.  We can write these generators as
\begin{equation}
L_{3} = \pmatrix{0 & -i & 0 \cr i & 0 & 0 \cr 0 & 0 & 0 }, \quad
Q_{1} = \pmatrix{0 & 0 & 0 \cr 0 & 0 & 0 \cr i & 0 & 0 }, \quad
Q_{2} = \pmatrix{0 & 0 & 0 \cr 0  & 0 & 0 \cr 0 & i & 0 } ,
\end{equation}
applicable to the three-dimensional space of $(x, y, z).$  They then
satisfy the closed set of commutation relations
\begin{equation}\label{eq300}
\left[Q_{1}, Q_{2}\right] = 0, \qquad
\left[L_{3}, Q_{1}\right] = iQ_{2}, \qquad
\left[L_{3}, Q_{2}\right] = -iQ_{1}.
\end{equation}
which becomes that of Eq.(\ref{eq200}) when $Q_{1}, Q_{2}, $ and
$L_{3}$ are replaced by $N_{1}, N_{2},$ and $J_{3}$ of Eq.(\ref{eq005})
respectively.  Indeed, this  cylindrical group is locally isomorphic
to Wigner's little group for massless particles.

Let us go back to the generators of Eq.(\ref{eq005}).  The role of
$J_{3}$ is well known.  It is generates rotations around the momentum and
corresponds to the helicity of the massless particle.
The $N_{1}$ and $N_{2}$ matrices take the form~\cite{bkn15}
\begin{equation}\label{eq6}
N_{1} = \pmatrix{0&0&-i&i\cr 0&0&0&0 \cr i&0&0&0 \cr i&0&0&0} , \qquad
N_{2} = \pmatrix{0&0&0&0 \cr 0&0&-i&i \cr 0&i&0&0 \cr 0&i&0&0} .
\end{equation}
The transformation matrix is
\begin{eqnarray}\label{eq101}
&{}& D(u,v) = \exp{\left\{-i\left(uN_{1} + vN_{2}\right)\right\}}
     \nonumber \\[1.0ex]
&{}&\hspace{20mm} = \pmatrix{1 & 0 & -u & u \cr 0 & 1 & -v & v \cr
u & v & 1 - (u^{2}+ v^{2})/2 & (u^{2} + v^{2})/2 \cr
u & v & -(u^{2} + v^{2})/2 & 1  + (u^{2} + v^{2})/2} .
\end{eqnarray}
If this matrix is applied to the electromagnetic wave propagating
along the $z$ direction,
\begin{equation}
A^{\mu}(z,t) = (A_{1}, A_{2}, A_{3}, A_{0}) e^{i\omega (z - t)} ,
\end{equation}
which satisfies the Lorentz condition $A_{3} = A_{0}$,
the $D(u,v)$ matrix can be reduced to~\cite{hks82}
\begin{equation}\label{eq109}
D(u,v) = \pmatrix{1&0&0&0 \cr 0&1&0&0\cr u&v&1&0\cr u&v&0&1} .
\end{equation}
If $A_{3} = A_{0}$, the four-vector $(A_{1}, A_{2}, A_{3}, A_{3})$
can be written as
\begin{equation}
(A_{1}, A_{2}, A_{3}, A_{0}) =
(A_{1}, A_{2}, 0, 0) + \lambda (0, 0, \omega, \omega) ,
\end{equation}
with $A_{3} = \lambda \omega$.  The four-vector $(0, 0, \omega, \omega)$
represents the four-momentum.  If the $D$ matrix of Eq.(\ref{eq109}) is
applied to the above four vector, the result is
\begin{equation}
(A_{1}, A_{2}, A_{3}, A_{0}) =
(A_{1}, A_{2}, 0, 0) + \lambda'(0, 0, \omega, \omega) ,
\end{equation}
with $\lambda' = \lambda + (1/\omega)\left(uA_{1} + vA_{3}\right)$.
Thus the $D$ matrix performs a gauge transformation when applied to
the electromagnetic wave propagating along the
$z$ direction~\cite{kiwi90jmp,hks82,bkn15}.

\section{Massless particle as a limiting case of massive particle}\label{o3e2}
From the generators of the Lorentz group, it is possible to construct
the four-by-four matrices for rotations around the $y$ axis and Lorentz
boosts along the $z$ axis as~\cite{bkn15}
\begin{equation}
 R(\theta) = \exp{\left(-i\theta J_{2}\right)}, \quad\mbox{and}\qquad
 B(\eta) = \exp{\left(-i\eta K_{3}\right)},
\end{equation}
respectively.  The Lorentz-boosted rotation matrix is $B(\eta) R(\theta) B(-\eta)$
which can be written as
\begin{equation}\label{eq600}
\pmatrix{\cos\theta  & 0 & (\sin\theta)\cosh\eta & -(\sin\theta)\sinh\eta \cr
         0 & 1 & 0  & 0 \cr
       -(\sin\theta)\cosh\eta & 0 & \cos\theta - (1 - \cos\theta)\sinh^2\eta &
        (1 - \cos\theta)(\cosh\eta) \sinh\eta \cr
       -(\sin\theta)\cosh\eta & 0 & -(1 - \cos\theta)(\cosh\eta)\sinh\eta &
        \cos\theta + (1 - \cos\theta)\cosh^{2}\eta} .
\end{equation}
While $\tanh\eta = v/c$, this boosted rotation matrix becomes a transformation
matrix for a massless particle when $\eta$ becomes infinite. On the other hand,
it the matrix is to be finite in this limit, the angle $\theta$ has to become
small.  Let us set
\begin{equation}\label{gamma}
\gamma = \frac{1}{2}\theta e^{\eta} .
\end{equation}
Then this four-by-four matrix becomes
\begin{equation}\label{eq601}
\pmatrix{1  & 0 & \gamma & - \gamma \cr
         0 & 1 & 0  & 0 \cr
       -\gamma & 0 & 1 - \gamma^2/2 &  \gamma^2/2  \cr
       -\gamma & 0 & -\gamma^2/2 & 1 + \gamma^2/2} .
\end{equation}
This is the Lorentz-boosted rotation matrix around the $y$ axis.  However,
we can rotate this $y$ axis around the $z$ axis by $\alpha$.  Then
the matrix becomes
\begin{equation}\label{eq603}
\pmatrix{1  & 0 & \gamma \cos\alpha & -\gamma \cos\alpha \cr
         0 & 1 &  \gamma \sin\alpha  & -\gamma\sin\alpha \cr
       -\gamma \cos\alpha & -\gamma \sin\alpha & 1 - \gamma^2/2 &  \gamma^2/2  \cr
       -\gamma \cos\alpha & -\gamma \sin\alpha & -\gamma^2/2 & 1 + \gamma^2/2} .
\end{equation}
this matrix becomes $D(u,v)$ of Eq.(\ref{eq101}), if we let
\begin{equation}\label{eq602}
u = -\gamma\cos\alpha, \quad\mbox{and}\quad  v = -\gamma\sin\alpha,
\end{equation}

\section{Spin-1/2 particles}\label{spinhalf}
Let us go back to the Lie algebra of the Lorentz group given in Eq.(\ref{eq002}).
It was noted that there are six four-by-four matrices satisfying nine
commutation relations.  It is possible to construct the same Lie algebra with
six two-by-two matrices~\cite{bkn15}.  They are
\begin{equation}
J_{i} = \frac{1}{2} \sigma_{i}, \quad\mbox{and}\quad
    K_{i} = \frac{i}{2} \sigma_{i} ,
\end{equation}
where $\sigma_{i}$ are the Pauli spin matrices.  While $J_{i}$ are Hermitian,
$K_{i}$ are not.  They are anti-Hermitian.  Since the Lie algebra of Eq.(\ref{eq002})
is Hermitian invariant, we can construct the same Lie algebra with
\begin{equation}
J_{i} = \frac{1}{2} \sigma_{i}, \quad\mbox{and}\quad
    \dot{K}_{i} = -\frac{i}{2} \sigma_{i} .
\end{equation}
This is the reason why the four-by-four Dirac matrices can explain both
the spin-1/2 particle and the anti-particle.

\par
Here again the $J_{i}$ matrices generate the rotation-like $SU(2)$ subgroup.
Here also we can consider the $E(2)$-like subgroup generated by
\begin{equation}
J_{3}, \qquad N_{1} = K_{1} - J_{2}, \qquad N_{2} = K_{2} + J_{1} .
\end{equation}
The $N_{1}$ and $N_{2}$ matrices take the form
\begin{equation}
N_{1} = \pmatrix{0 & i \cr 0 & 0}, \qquad N_{2} = \pmatrix{0 & 1 \cr 0 & 0} .
\end{equation}

On the other hand, in the ``dotted'' representation,
\begin{equation}
\dot{N}_{1} = \pmatrix{0&0 \cr -i & 0} , \qquad
\dot{N}_{2} = \pmatrix{0&0\cr 1&0}.
\end{equation}
There are therefore two different $D$ matrices:
\begin{equation}
D(u,v) = exp{\left\{-\left(iuN_{1} + ivN_{2}\right)\right\}}
 = \pmatrix{1 & u + iv \cr 0&1} ,
\end{equation}
and
\begin{equation}
\dot{D}(u,v) = exp{\left\{-\left(iu\dot{N}_{1} + iv\dot{N}_{2}\right)\right\}}
  = \pmatrix{1 & 0 \cr u - iv & 1} .
\end{equation}
These are the gauge transformation matrices applicable to  massless spin-1/2
particles~\cite{hks82,bkn15}.

We are familiar with the notation
\begin{equation}
\chi_{+} = \pmatrix{1 \cr 0}, \quad\mbox{and}\quad \chi_{-} = \pmatrix{0 \cr 1},
\end{equation}
In addition, we need two additional spinors
\begin{equation}\label{eq701}
\dot{\chi}_{+} = \pmatrix{1 \cr 0}, \quad\mbox{and}\quad
   \dot{\chi}_{-} = \pmatrix{0 \cr 1}.
\end{equation}
Here are talking about the Dirac equation for with four-component spinors.

The spinors $\chi_{+}$ and $\dot{\chi}_{-}$ are gauge-invariant since
\begin{equation}
         D(u,v)\chi_{+} =\chi_{+}, \quad\mbox{and}\quad
         \dot{D}(u,v)\dot{\chi}_{-} = \dot{\chi}_{-} .
\end{equation}
As for $\chi_{-}$ and $\dot{\chi}{+}$,
\begin{eqnarray}\label{eq702}
&{}& D(u,v) \chi_{-} = \chi_{-} + (u - iv)\chi_{+} , \nonumber \\[2mm]
&{}& \dot{D}(u,v) \dot{\chi}_{+} =
\dot{\chi}_{+} + (u + iv)\dot{\chi}_{-} .
\end{eqnarray}
They are not gauge invariant.  Thus, we can conclude that the polarization
of massless neutrinos is a consequence of gauge invariance.

In this two-by-two representation, the Lorentz boost along the positive
direction is
\begin{equation}
     B(\eta) = \pmatrix{e^{\eta/2} & 0 \cr 0 & e^{-\eta/2}},
\end{equation}
the rotation around the $y$ axis is
\begin{equation}
R(\theta) = \pmatrix{\cos(\theta/2) & -\sin(\theta/2) \cr
                  \sin(\theta/2) & \cos(\theta/2)} .
\end{equation}
Then, the boosted rotation matrix is
\begin{equation}\label{eq607}
B(\eta) R(\theta) B(-\eta) =
         \pmatrix{\cos(\theta/2) & - e^{\eta}\sin(\theta/2) \cr
                  e^{-\eta}\sin(\theta/2) & \cos(\theta/2)} .
\end{equation}

If $\eta$ becomes very large, and this matrix is to remain finite,
$\theta$ has to become very small, and this expression becomes
\begin{equation}
 \pmatrix{1 - \gamma^2 e^{-2\eta}/2 & - \gamma \cr
                  \gamma e^{-2\eta} & 1 - \gamma^2 e^{-2\eta}/2} .
\end{equation}
with $ \gamma = \frac{1}{2}\theta e^{\eta}$ as  given in Eq.(\ref{gamma}).
This expression becomes
\begin{equation}
D(\gamma) = \pmatrix{1 & - \gamma \cr 0 & 1} .
\end{equation}
In this two-by-two representation, the rotation around the $z$ axis is
\begin{equation}
Z(\phi) = \pmatrix{e^{i\phi/2} & 0 \cr 0 & e^{-i\phi/2}}, \qquad
\end{equation}
respectively.  Thus
\begin{equation}
  D(u,v) = Z(\phi + 180) D(\gamma) Z^{-1}(\phi + 180) ,
\end{equation}
 which becomes
   \begin{equation}
       D(u, v) = \pmatrix{1 & u + iv \cr 0 & 1} .
   \end{equation}
with $u$ and $v$ given in Eq.(\ref{eq602}).

We have studied how the massive neutrino with its O(3)-like symmetry
to become the massless neutrino in the massless limit.  The boost
parameter $\eta$ can be derived from the mass and momentum of the
massive neutrino from
\begin{equation}
\tanh\eta = \frac{p}{\sqrt{m^2 + p^2}},
\end{equation}
where $m$ and $p$ are the mass and the momentum of the neutrino
respectively.  For small values of $m/p$,
\begin{equation}
          e^{\eta} = \frac{\sqrt{2} p}{m} ,
\end{equation}
which becomes large when $m$ becomes very small.

The question is then the reverse process~\cite{kmn16}. Start from the massless
neutrino with its gauge degree of freedom.   What happens
when the particle gains its mass? In that case, the particle can
be brought to its rest frame?  The matrix of Eq.(\ref{eq607}) is one
way to bring it the rest frame, and $\gamma$ becomes $\sin\theta$.
However, what happens when $\gamma$ is larger than one?  This is
a  challenging future question.

\end{document}